# Strong Second Harmonic Generation from Bilayer Graphene with Symmetry Breaking by Redox-Governed Charge Doping

*Mingwen Zhang[1, §], Nannan Han[2, §], Jing Wang[1], Zhihong Zhang[3], Kaihui Liu[3], Zhipei Sun[4],*

*Jianlin Zhao[1], Xuetao Gan[1,] \**

\*Email: xuetaogan@nwpu.edu.cn

1. Key Laboratory of Light Field Manipulation and Information Acquisition, Ministry of Industry and Information Technology, and Shaanxi Key Laboratory of Optical Information Technology, School of Physical Science and Technology, Northwestern Polytechnical University, Xi'an 710129, China

2. Frontiers Science Center for Flexible Electronics, Xi'an Institute of Flexible Electronics (IFE) and Xi'an Institute of Biomedical Materials & Engineering, Northwestern Polytechnical University, Xi'an 710129, China

3. State Key Lab for Mesoscopic Physics and Frontiers Science Center for Nano-optoelectronics, Collaborative Innovation Center of Quantum Matter, School of Physics, Peking University, Beijing 100871, China

4. Department of Electronics and Nanoengineering and QTF Centre of Excellence, Aalto University, Aalto FI-00076, Finland.

ABSTRACT

Missing second-order nonlinearity in centrosymmetric graphene overshadows its intriguing optical attribute. Here, we report redox-governed charge doping could effectively break the centrosymmetry of bilayer graphene (BLG), enabling a strong second harmonic generation (SHG) with a strength close to that of the well-known monolayer $MoS_2$. Verified from control experiments with *in situ* electrical current annealing and electrically gate-controlled SHG, the required centrosymmetry breaking of the emerging SHG arises from the charge-doping on the bottom layer of BLG by the oxygen/water redox couple. Our results not only reveal that charge doping is an effective way to break the inversion symmetry of BLG despite its strong interlayer coupling but also indicate that SHG spectroscopy is a valid technique to probe molecular doping on two-dimensional materials.

Responses of a material to an applied optical field $E(t)$ are governed by the polarization $P(t)=\varepsilon_0[\chi^{(1)}E(t)+\chi^{(2)}E(t)E(t)+\chi^{(3)}E(t)E(t)E(t)+...]$, where $\chi^{(1)}$, $\chi^{(2)}$, and $\chi^{(3)}$ are the linear and second- and third-order nonlinear susceptibilities. Benefiting from the unique linear dispersion and massless Dirac fermions,[1] graphene represents an attractive optical material for both linear and nonlinear optics. In the linear optical regime, a single layer of graphene has a universal absorbance (determined by the imaginary part of $\chi^{(1)}$) of 2.3% from the visible to the mid-infrared.[2] In the nonlinear optical regime, graphene permits enhanced nonlinear optical effects over a wide spectral range due to the interband optical transition and the low density of states around the Dirac point.[3] This enables relatively high $\chi^{(3)}$ ($10^{-7}$ esu),[4] giving rise to strong third harmonic generation (THG)[5] and four-wave-mixing[4] for high-contrast optical microscopy and visible light source. In addition, graphene's single-atom thickness and linear dispersion with a high Fermi velocity allow effective tuning of its Fermi energy through electrostatic gating,[6] promising electrically controlled optical responses.[7] For example, the prohibition of linear optical absorption[3,8] and enhanced $\chi^{(3)}$ in graphene by 1 order of magnitude by controlling the Fermi level were realized.[9,10]

Unfortunately, the absence of second-order nonlinearity in the centrosymmetric graphene overshadows its decent optical responses, considering $\chi^{(2)}$ is more attractive than $\chi^{(3)}$ due to the much higher coefficient. Numerous efforts have thus far been made to pursue $\chi^{(2)}$ in graphene. In monolayer graphene (MLG), $\chi^{(2)}$-supported second harmonic generation (SHG) is possible when the surface dipole[11] or the radiation wave vector[12] is taken into account. With the assistance of an in-plane photon wavevector, SHG with an electric-quadrupole response was realized in highly doped graphene.[13] In bilayer graphene (BLG), it is theoretically proposed that an in- or out-of-plane electric field can be applied to break the inversion symmetry,[14,15] giving rise to weak $\chi^{(2)}$-supported SHG. In accidental areas of exfoliated trilayer graphene, if the stacking order is ABA (Bernal-type), electric-dipole-enabled SHG is allowed because the $D_{3h}$ symmetry has no inversion center.[16] By artificially stacking two graphene monolayers with precisely controlled twisted angles, it is also possible to obtain a two-dimensional system without centrosymmetry for SHG.[17] These symmetry breakings in graphene for SHG are mostly achieved by an external electric field or intentional stacking.

Here, we report a simple strategy to realize remarkable $\chi^{(2)}$ in naturally AB-stacked BLG via redox-governed charge doping. With annealing treatment of the pristine BLG, oxygen ($O_2$) and water ($H_2O$) molecules are adsorbed at the interface of BLG and hydrophilic $SiO_2$, which induces an interlayer built-in electric field and inversion symmetry breaking. It is confirmed by control experiments with *in situ* electrical current annealing and electrically controlled SHG intensity, combined with polarization-resolved SHG measurements. The emerging SHG signal from the charge-doped BLG is comparable to that measured from the monolayer molybdenum disulfide ($MoS_2$), indicating that redox-governed charge doping is an effective way to break the inversion symmetry of BLG in spite of its strong interlayer coupling. Our results also reveal that SHG spectroscopy is a more reliable technique to determine the molecular induced charge doping on two-dimensional materials than the widely employed Raman, photoluminescence, and pump-probe spectroscopies.

We fabricate the BLG sample via standard mechanical exfoliation onto a heavily doped silicon substrate covered with a silicon oxide layer of 285 nm thickness. Figure 1a shows the optical microscope image of an exfoliated graphene sample consisting of both MLG and BLG, which are confirmed by the optical contrast method[18] and Raman spectroscopy[19] (Figure S1).

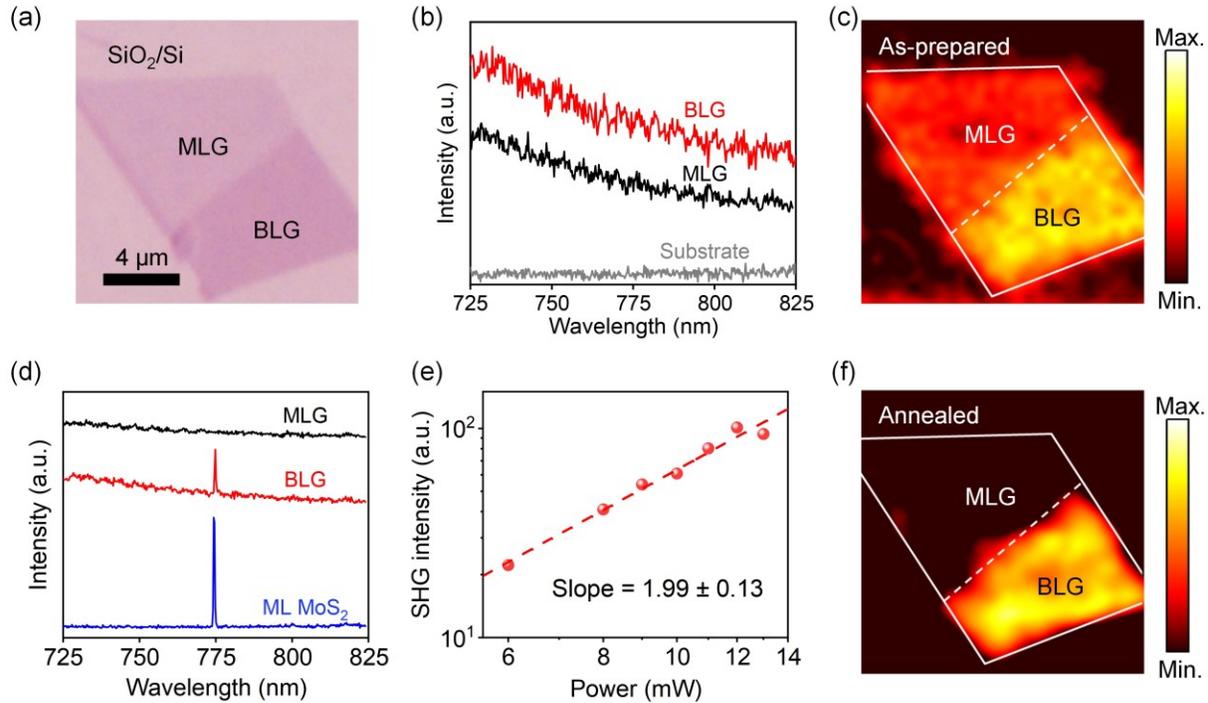

**Figure 1.** (a) Optical microscope image of the as-prepared graphene sample. (b) Frequency up-conversion spectra from different regions of the as-prepared graphene sample. (c) Spatial mapping of NPL signal from the as-prepared graphene sample. (d) Frequency up-conversion spectra from the MLG and BLG regions of the annealed graphene sample. SHG signal from a ML $MoS_2$ is shown as well for comparison. (e) The log-log plot of the SHG intensity from the BLG as a function of incident laser power, showing a linear fit with the extracted slope of $1.99\pm0.13$. (f) SHG spatial mapping of the annealed graphene sample.

We first characterize the as-prepared graphene sample on a home-built multiphoton nonlinear optical microscopy system. A picosecond pulsed laser at the wavelength of 1550 nm is chosen as the fundamental pump radiation. More details about the optical setup can be seen in Figure S2. Figure 1b displays the acquired frequency up-conversion optical spectra from the regions of MLG and BLG shown in Figure 1a. No SHG but only broadband nonlinear photoluminescence (NPL) can be observed from MLG and BLG.[20,21] Then the graphene sample is mounted on a two-dimensional piezoactuated stage for spatial mapping of the NPL signal, as shown in Figure 1c. Clear regions with different intensities are distinguished from the MLG and BLG. The thicker BLG has stronger NPL than the MLG, and both of them have uniform distributions.

As well studied in the past decade, with a high-temperature vacuum annealing, graphene becomes more hole-doped after exposure to air due to the oxygen reduction reaction (ORR).[22–25] To introduce the $O_2/H_2O$ redox couple in BLG-$SiO_2$, we then anneal the as-prepared graphene sample in a vacuum furnace at 200 °C for 30 min and then expose it to an air ambient. This annealing condition introduces no additional defects in graphene,[26] which is verified by the Raman spectra without the defect-related D band (see Figure S3). After that, the frequency up-conversion signals are acquired again, as shown in Figure 1d. From the region of BLG, there is a strong peak at the wavelength of 775 nm over the broadband NPL background (red curve). We note here that the defects caused by high laser power in the SHG test can be observed, but no SHG was observed at the corresponding position, as shown in Figure S3. That is, the defects cannot give rise to SHG. To examine the origin of this peak, the dependence of its intensity on the pump power is measured, as shown in Figure 1e. A linear slope of $1.99\pm0.13$ is obtained in the log-log scale, which is consistent with the SHG process that two photons of the fundamental pump laser convert into one photon of the SHG signal. On the contrary, from the region of MLG, only broadband NPL is observed.

To evaluate the strength of the obtained SHG from the BLG, we also measure the SHG from a monolayer $MoS_2$ (ML $MoS_2$) with the same pump conditions, as shown in Figure 1d. The intensity of the emerging SHG signal from the BLG is about 40% of that from ML $MoS_2$. Though the SHG from the BLG is weaker than that from the ML $MoS_2$, their similar strengths indicate that the doping-induced $\chi^{(2)}$ from BLG is comparable to that of the widely studied ML $MoS_2$, which is well recognized with large $\chi^{(2)}$.[27] In other words, the charge doping breaks the inversion symmetry of BLG effectively.

The charge doping enabled SHG from the BLG is further illustrated by implementing its spatial mapping. The mapping result after annealing is displayed in Figure 1f after subtracting the NPL background, showing a clear and continuous region with an obvious SHG signal from the BLG. No SHG is observed from the MLG. It also implies the uniformity of charge doping.

To verify that the observed remarkable SHG from the BLG is induced by the redox-governed charge doping of $O_2$ and $H_2O$ molecules, we carry out control experiments by measuring the variations of SHG signal from the BLG when it is *in situ* annealed by an electrical current-induced Joule heat. As shown in Figure 2a, a BLG field effect transistor (FET) is fabricated. Two gold (Au) pads are contacted at two sides of the graphene sheet as the electrical drain and source electrodes. The highly doped silicon substrate functions as the global back gate. In the optical measurements, the fabricated BLG FET is loaded in a gas-flowed chamber, which has a top glass window to allow the transmissions of the pump laser and SHG signal, as schematically shown in Figure 2b. When a large electrical voltage ($V_{ds}$) is applied across the drain and source electrodes, the BLG sheet could generate considerable Joule heat to anneal itself,[28] which allows the *in situ* monitoring of SHG variations from the BLG.

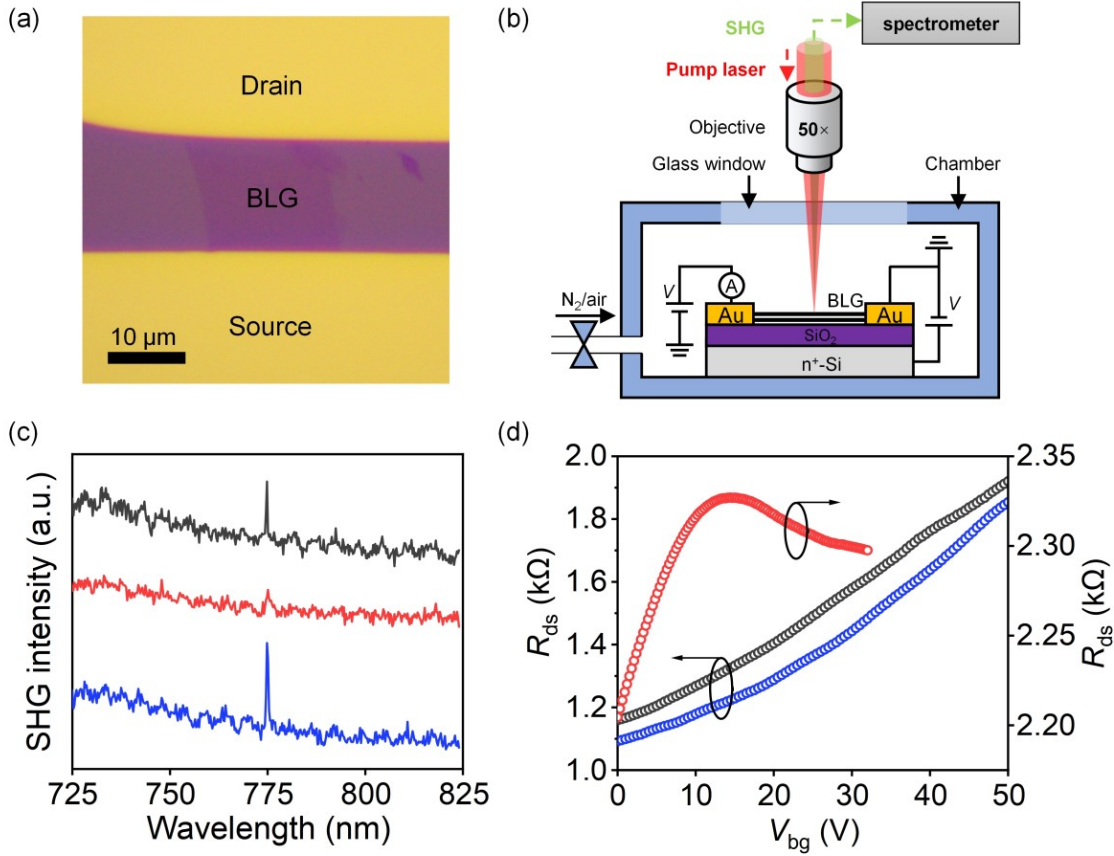

**Figure 2.** (a) Optical microscope image of the BLG FET. (b) Schematic diagram (side view) of the experimental setup for *in situ* monitoring variations of BLG's SHG during electrical current annealing. (c) SHG spectra of the BLG acquired before the current annealing (blue curve), after the current annealing in $N_2$ environment (red curve), and after the re-exposure in ambient air (black curve). (d) Electrical transport properties of BLG FET (electrical resistance versus gate voltage)

acquired before the current annealing (blue curve), after the current annealing in $N_2$ environment (red curve), and after the re-exposure in ambient air (black curve). The $R_{ds}$ range of the red curve is enlarged and displayed on the right y-axis for clear comparison with the charge neutral points of other curves.

We characterize the variations of BLG's SHG in three steps. In the first step, we measure the SHG from the as-fabricated BLG FET before the electrical annealing. The result is shown as the blue curve in Figure 2c. Here, during the fabrication of the FET device, at the final process, we anneal it in a vacuum furnace at 200 °C for 30 min and then expose it to ambient air, which is expected to introduce adsorption of $O_2$ and $H_2O$ molecules. Similar to the result in Figure 1b, a remarkable SHG signal is observed from the BLG.

In the second step, an electrical bias $V_{ds}$ of 12 V is applied on the FET, and nitrogen ($N_2$) gas is flowed through the chamber simultaneously to establish an inert environment. The electrical current across the BLG channel becomes larger than 7 mA. This translates into a large current density of about $10^8$ A/cm$^2$. The corresponding Joule heat is large enough to remove the adsorbed molecules on graphene.[28] After electrical heating for 2 min, the $V_{ds}$ is switched off, while the $N_2$ environment is maintained. The SHG spectrum of the BLG device is measured again, as shown in the red curve of Figure 2c. With the same SHG measurement condition as that used for the blue curve in Figure 2c, the signal peak at the wavelength of 775 nm is almost unobservable. The SHG intensity is reduced to 20% of that before electrical annealing, which could be attributed to the removal of $O_2$ and $H_2O$ molecules with the high-temperature annealing by the current-induced Joule heat, as discussed above. However, when the sample is cooled to room temperature in the $N_2$ environment, the $N_2$ molecules induce much weaker charge doping on the graphene sheet, which cannot break the intrinsic inversion symmetry of BLG.[24] Consequently, no remarkable SHG is expected.

In the third measurement step, after cutting off the $N_2$ flow, the chamber is opened to expose the BLG sheet in the air ambient. With the constant SHG measurement condition as that in the above two steps, the SHG signal from the BLG is acquired, as shown in the black curve of Figure 2c. A strong signal peak presents at the wavelength of 775 nm, indicating the emerging SHG induced by the readsorbed $O_2$ and $H_2O$ molecules.

As discussed above, the emerging SHG from the BLG is given by the inversion symmetry breaking induced by the redox-governed charge doping on graphene which could also be verified by measuring the electrical transport properties of the fabricated BLG FET.[22] Figure 2d shows the variations of electrical resistance ($R_{ds}$) of the BLG sheet under different gate voltages ($V_{bg}$) acquired in the three steps defined in the above SHG measurements. In the first step, $R_{ds}$ of the as-fabricated BLG FET before the current annealing is measured and shown as the blue curve in Figure 2d. The charge neutral point is larger than $V_{bg}=50$ V, indicating heavy hole-doping in BLG after vacuum annealing and exposure to ambient air. The hole concentration of the BLG is calculated as larger than $n\sim 3.6\times 10^{12}$ cm$^{-2}$ using the equation $n=C_{bg}V_{bg}/e$, as in ref. 22.

In the second step, the transfer curve of the BLG FET is measured again in the $N_2$ atmosphere after the electrical current annealing for 2 min, as shown in the red curve of Figure 2d. The charge neutral point shifts to less than $V_{bg}=15$ V, indicating that the BLG sample is only slightly p-doped where the hole concentration is calculated as less than $n\sim1.08\times10^{12}$ cm$^{-2}$. That is, the $O_2$ and $H_2O$ molecules are removed by the Joule heat during the current annealing. $N_2$ in the chamber has hardly any doping effect on BLG.

In the third step, after the $N_2$ flow is cut off and the chamber is opened, the measurement of the transfer curve is carried out. As shown in the black curve of Figure 2d, the charge neutral point of the FET transfer curve shifts back to be larger than 50 V, which means $O_2$ and $H_2O$ molecules in the air are readsorbed. It is concluded that, when the annealed BLG is exposed to air, the $O_2$ adsorption induces hole doping as well as the emerging strong SHG.

Here, we explain that the hole doping of graphene is governed by the ORR involving oxygen and water,

$$O_2 + 4H^+ + 4e^- \leftrightarrow 2H_2O, \tag{1}$$

where electrons are taken from graphene. As well studied in ref. 25, the graphene-$SiO_2$ interface will be hydroxylated by thermal activation, which renders the $SiO_2$ surface hydrophilic enough to bind a certain amount of $H_2O$ molecules. When the annealed graphene-$SiO_2$ is exposed to the ambient air, the $O_2$ molecules in the air diffuse through the interface of the graphene-$SiO_2$, as illustrated in Figure 3a. The ORR mainly takes electrons from the bottom layer of BLG, resulting in the hole doping of graphene.

The above analysis indicates that the charge transfer (CT) between the $O_2/H_2O$ redox couples and the bottom layer of BLG gives rise to the unbalanced charge concentration between the top and bottom monolayers. Consequently, a charge doping induced interlayer electric field ($E_{induced}$) is present between the top and bottom layers,[29] which is approximately perpendicular to the graphene surface, as schematically shown in Figure 3b. The inversion symmetry in BLG is broken due to the interlayer electric field and the inequivalent potentials of the two layers.[30,31] For pristine BLG with inversion symmetry, which belongs to the $D_{3d}$ point group,[32] all $\chi^{(2)}$ elements are zero and the SHG vanishes. However, with the unbalanced charge concentration and inequivalent potentials between the top and bottom graphene layer, the inversion symmetry is broken, which converts it into the $C_{3v}$ point group.[31,33] As a result, nonvanishing $\chi^{(2)}$ elements emerge and the SHG is nonzero.

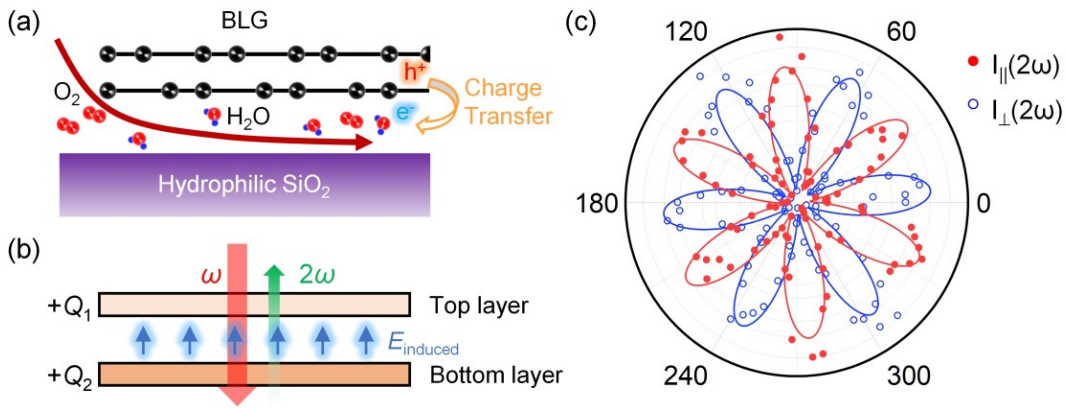

**Figure 3.** (a) Scheme for major diffusion routes of $O_2$ responsible for ORR. (b) Schematic diagram of the doping-induced electric field ($E_{induced}$) in BLG. Different colors in the top and bottom graphene layers indicate different charge doping levels induced by ORR. (c) Polarization-resolved patterns of SHG intensity from the BLG sample. The circles and dots are the measured SHG signal, and the curves are the function fittings.

The polarization dependences of SHG from the charge-doped BLG are further studied. Figure 3b shows the polarization-resolved SHG patterns where the polarizations of the excitation and signal beams are parallel and perpendicular, respectively. Note that the dependence of the SHG electric field on crystallographic orientation exhibits 3-fold rotational symmetry under normal incidence, following the lattice symmetry of the $C_{3v}$ point group. The parallel and perpendicular components of the SHG intensities could be described as

$$I_\parallel(2\omega) \propto (d_{22} \sin(3\theta))^2 \qquad (2)$$

and

$$I_\perp(2\omega) \propto (-d_{22} \cos(3\theta))^2 \qquad (3)$$

respectively (see Supporting Information for detailed calculation process). Here, $d_{22}$ is the responsible $\chi^{(2)}$ susceptibility tensor element[34] and $\theta$ is the angle between the armchair direction and the excitation polarization.[35] Equations (2) and (3) are used to fit the SHG polarization dependence, as shown in the solid lines in Figure 3b, which confirms that doped BLG belongs to the $C_{3v}$ point group. That is, the ORR-induced interlayer built-in electric field in BLG effectively breaks its inversion symmetry and gives rise to the strong SHG.

To confirm the locations of dopants and the direction of the built-in electric field, this ORR-induced SHG has been electrically tuned with a back-gate of the BLG FET, as shown in Figure 4. The top inset is the schematic illustration of the fabricated FET device. The bottom inset shows the optical microscope image of the BLG FET. As $V_{bg}$ increases from -100 V to 100 V, the upward electric field induced by $V_{bg}$ increases, which has the same direction as that of $E_{induced}$ shown in Figure 3b. As a result, the total electric field across the BLG increases, which improves the SHG intensity by about 5 times. It makes sense only when the direction of the $E_{induced}$ is upward, which also proves that the dopant is located below graphene.

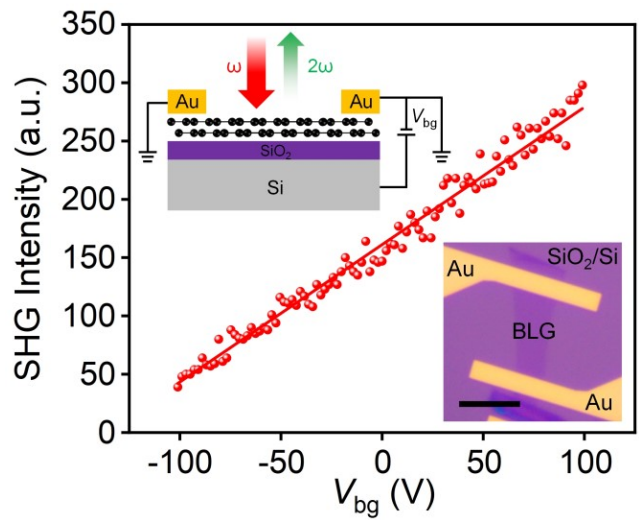

**Figure 4.** SHG intensity in BLG as a function of $V_{bg}$. Top inset: schematic illustration of the BLG FET. Bottom inset: optical microscope image of the BLG FET. Scale bar: 10 μm.

For MLG, the ORR could also cause hole doping and symmetry breaking in the graphene layer which can be described by the point group $C_{6v}$.[31] However, since electric-dipole-allowed SHG is not sensitive to the presence of a $C_n$ rotation symmetry operation with n > 3, no SHG signal is expected here.[11] In the case of BLG, the interlayer electric field is formed due to the inequivalent potentials and strong interlayer coupling between the top and bottom monolayers, resulting in the $C_{3v}$ symmetry and the emergence of SHG. We note here that the interlayer coupling plays an important role in SHG, which has been demonstrated in the field-induced SHG of the AB-stacked $MoS_2$ bilayer[36] where the $\chi^{(2)}$ is reduced dramatically (> 3 orders of magnitude) by artificially doubling the interlayer distance.

Besides SHG, the change of THG intensity upon charge doping by ORR has also been examined in our experiments, which is also shown in Figure S4. The charge doping by ORR on the graphene has little influence on the THG intensity of BLG since the doping is not high enough to shift the Fermi level to alter the $\chi^{(3)}$ of graphene.[9,10]

Following the initial experiments of BLG samples prepared by mechanical exfoliation, we also test the SHG characteristics of a BLG sample synthesized by the copper-catalyzed chemical vapor deposition (CVD) technique. The results are shown in Figure S5. SHG appeared in CVD-grown BLG after annealing and being exposed to air, but there was no SHG in freshly prepared samples. This shows that the charge doping induced SHG is also valid on the CVD-grown BLG samples, which provides a way to prepare large-scale BLG with strong second-order nonlinearity for nonlinear optical devices.

We presented a novel emerging strong SHG in AB-stacked BLG induced by redox-governed charge doping, which has a similar strength as that from a monolayer $MoS_2$. Annealing treatment promotes the ORR, leading to charge doping and inversion symmetry breaking in BLG. It is experimentally proved by measuring SHG variations with *in situ* electrical current annealing. The CT between the $O_2/H_2O$ redox couple and BLG gives rise to the formation of an interlayer built-in electric field between the top and bottom monolayers due to the inequivalent charge distributions. With that, the BLG changes from $D_{3d}$ to $C_{3v}$ point group, which is further confirmed by the polarization-resolved SHG and electrically gate-controlled SHG.

Our results reveal that the charge doping on BLG by ORR is significant enough to induce an interlayer electric field comparable to that constructed by BLG's intrinsic interlayer interaction, giving rise to inversion symmetry breaking. It provides a strategy to engineer nonlinear optical responses of graphene or other two-dimensional materials. On the other hand, the interface is important for two-dimensional materials. Though spectroscopy techniques of Raman, photoluminescence, and pump-probe are normally utilized to determine their interface states, our results indicate the SHG spectroscopy is more reliable and straightforward by characterizing whether SHG emerges or not in the centrosymmetric materials.

ASSOCIATED CONTENT

**Supporting Information.**

Determination of the thicknesses of graphene samples; Multiphoton nonlinear optical microscopy system; Raman spectroscopy characterization and SHG of defects in BLG; Change of THG intensity upon the charge doping by molecular adsorption; Calculation of the parallel and perpendicular components of the SHG intensities; SHG in CVD-grown BLG.

AUTHOR INFORMATION

**Corresponding Author**

*Xuetao Gan - Key Laboratory of Light Field Manipulation and Information Acquisition, Ministry of Industry and Information Technology, and Shaanxi Key Laboratory of Optical Information Technology, School of Physical Science and Technology, Northwestern Polytechnical University, Xi'an 710129, China

Email: xuetaogan@nwpu.edu.cn

**Author Contributions**

The manuscript was written through contributions of all authors. All authors have given approval to the final version of the manuscript. §M.Z. and N.H. contributed equally to this work.

**Notes**

The authors declare no competing financial interest.

ACKNOWLEDGMENT

This work is supported by the National Key R&D Program of China (Grant Nos. 2018YFA0307200 and 2017YFA0303800), the National Natural Science Foundation of China (Grant Nos. 61775183, 11634010, 61905196), the Key R&D Program of Shaanxi Province (Grant Nos. 2020JZ-10), Key R&D Program of Guangdong Province (2020B010189001, 2019B010931001, and 2018B030327001), the Fundamental Research Funds for the Central Universities (Grant Nos. 3102017jc01001, 3102019JC008, 310201911cx032).**REFERENCES**

(1) Novoselov, K. S.; Geim, A. K.; Morozov, S. V.; Jiang, D.; Katsnelson, M. I.; Grigorieva, I. V.; Dubonos, S. V.; Firsov, A. A. Two-Dimensional Gas of Massless Dirac Fermions in Graphene. *Nature* **2005**, *438* (7065), 197–200 DOI: 10.1038/nature04233.

(2) Nair, R. R.; Blake, P.; Grigorenko, A. N.; Novoselov, K. S.; Booth, T. J.; Stauber, T.; Peres, N. M. R.; Geim, A. K. Fine Structure Constant Defines Visual Transparency of Graphene. *Science (80-. ).* **2008**, *320* (5881), 1308 DOI: 10.1126/science.1156965.

(3) Wang, F.; Zhang, Y.; Tian, C.; Girit, C.; Zettl, A.; Crommie, M.; Shen, Y. R. Gate-Variable Optical Transitions in Graphene. *Science (80-. ).* **2008**, *320* (5873), 206–209 DOI: 10.1126/science.1152793.18

# Strong Second Harmonic Generation from Bilayer Graphene with Symmetry Breaking by Redox-Governed Charge Doping

*Mingwen Zhang*[1, §], *Nannan Han*[2, §], *Jing Wang*[1], *Zhihong Zhang*[3], *Kaihui Liu*[3], *Zhipei Sun*[4], *Jianlin Zhao*[1], *Xuetao Gan*[1,][*]

*Email: xuetaogan@nwpu.edu.cn

1. Key Laboratory of Light Field Manipulation and Information Acquisition, Ministry of Industry and Information Technology, and Shaanxi Key Laboratory of Optical Information Technology, School of Physical Science and Technology, Northwestern Polytechnical University, Xi'an 710129, China

2. Frontiers Science Center for Flexible Electronics, Xi'an Institute of Flexible Electronics (IFE) and Xi'an Institute of Biomedical Materials & Engineering, Northwestern Polytechnical University, Xi'an 710129, China

3. State Key Lab for Mesoscopic Physics and Frontiers Science Center for Nano-optoelectronics, Collaborative Innovation Center of Quantum Matter, School of Physics, Peking University, Beijing 100871, China

4. Department of Electronics and Nanoengineering and QTF Centre of Excellence, Aalto University, Aalto FI-00076, Finland.

## 1. Determination of the thicknesses of graphene samples

The graphene flakes are first distinguished by optical microscope using optical contrast method. Then we use Raman spectroscopy to determine the specific layer numbers of the graphene samples. All Raman spectra are recorded at room temperature in the air using the confocal WiTec Alpha 300R Raman Microscope. The wavelength of the excitation laser is 532 nm and the power is 1 mW. Figure S1a shows the Raman spectra of the mechanically exfoliated monolayer and bilayer graphene (MLG and BLG) in Figure 1a, while Figure S1b shows the Raman spectra of the transferred CVD-grown samples. The 2D band for BLG is fitted using four Lorentzian peaks. All of them have the same FWHM of 24 $cm^{-1}$, which is consistent with the Raman properties of AB-stacked BLG reported in the literature.[1] No D band at ~1350 $cm^{-1}$ can be seen in the Raman spectra, which means that the sample is of high quality.[2]

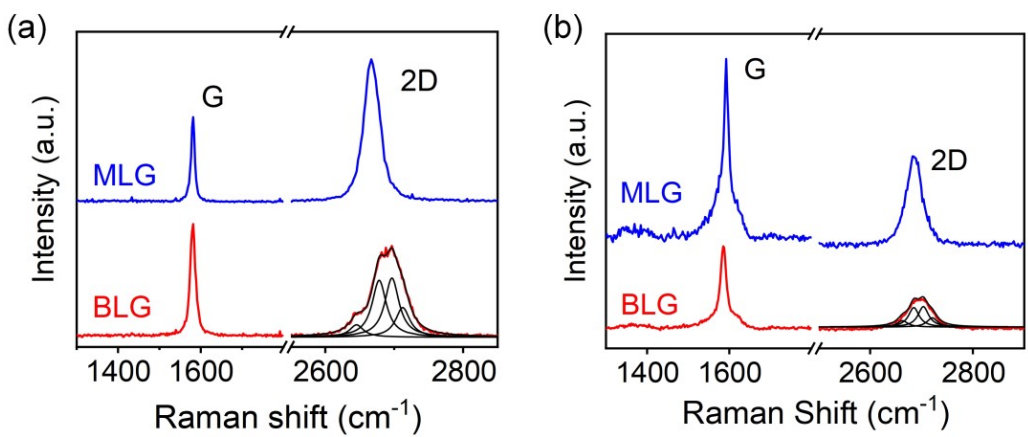

**Figure S1** Raman spectra of MLG and BLG prepared by mechanical exfoliation (a) and CVD-grown (b).

## 2. Multiphoton Nonlinear Optical Microscopy System

SHG measurements of the graphene samples are implemented in a home-built multiphoton nonlinear optical microscopy system with the reflection geometry, as illustrated in Figure S2. The fundamental pump from a picosecond pulsed laser at the wavelength of 1550 nm is linearly polarized via a polarizer. Then it is reflected by a dichroic mirror and focused on the sample through a 50× objective lens with a numerical aperture (NA) of 0.75. The generated SHG radiation is then back collected by the same objective and passes through the dichroic mirror, which is subsequently coupled to a spectrometer mounted with a cooled silicon CCD camera for spectra measurements.

For SHG spatial mapping, samples are placed on a piezo-actuated stage with a moving step of 100 nm for the in-plane scanning, while the excitation and collection light-spots are fixed. To study the polarization dependence of the SHG radiation, another polarizer is placed in the signal collection path, whose direction is rotated correspondingly to the pump polarization to collect the parallel or perpendicular components of the SHG signal. We note here that the polarizer is not placed in the signal path for spectra measurements.

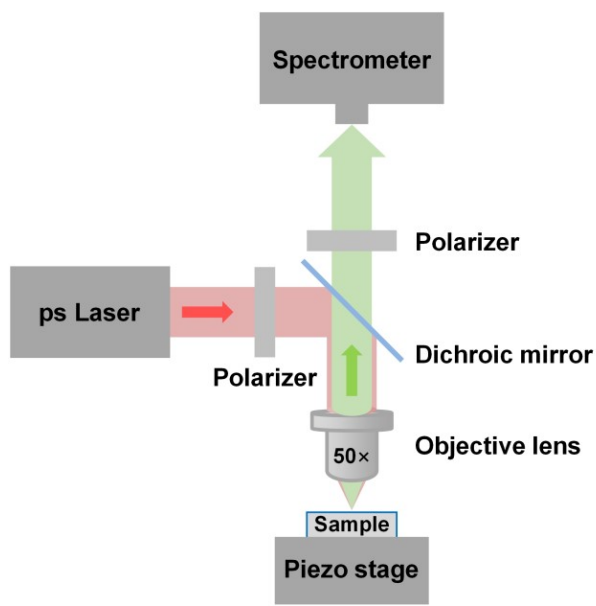

**Figure S2** Schematic illustration of multiphoton nonlinear optical microscopy system.

## 3. Raman spectroscopy characterization and SHG of defects in BLG

The Raman spectrum after annealing treatment is shown as the red curve in Figure S3a, where no defect-related D band is observed. However, as we increase the pump power in the SHG test, we observed the defects caused by high laser power by Raman spectroscopy, as shown in the blue curve in Figure S3a. The corresponding SHG results are shown in Figure S3b (the blue curve), where no SHG was observed at the defect position. The inset shows the optical microscope image of the sample in Figure S3c.

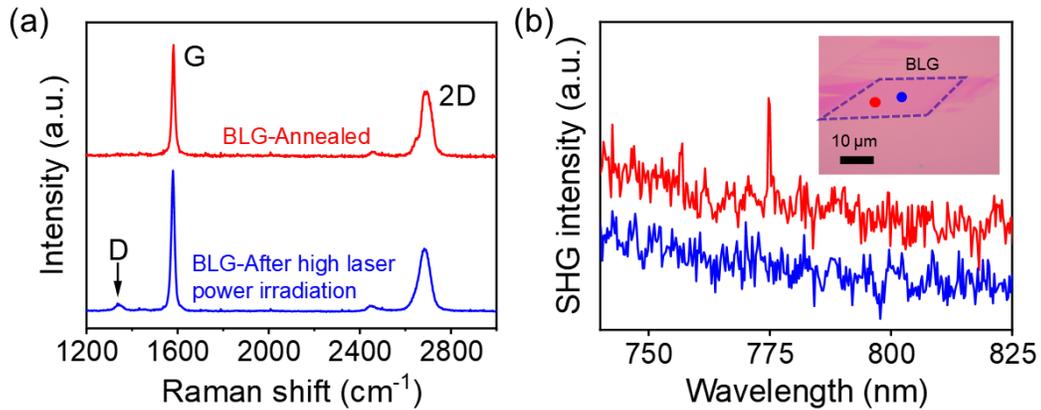

**Figure S3** Raman spectra (a) and SHG spectra (b) of annealed BLG and the BLG after high laser power irradiation. Inset in (b): Optical microscope image of the BLG sample. The red and blue dots indicate the corresponding position of the acquired spectra with the same colors in (a) and (b).

## 4. Change of THG intensity upon the charge doping by molecular adsorption

The wavelength of the excitation laser source used in the THG measurement is tuned to 1620 nm, so the THG wavelength is 540 nm. After vacuum annealing and exposing the BLG to air, the $O_2$ and $H_2O$ molecules in the air are adsorbed on graphene. As shown in Figure S5, the THG intensities of both MLG and BLG have no significant changes before and after annealing. This is due to the low doping concentration introduced by molecular adsorption. According to the transfer curve measured in the *in-situ* electrical annealing experiment (see Figure 2d of the main text) and the calculation method in the literature,[3] the doping concentration introduced by molecular adsorption in graphene is less than $10^{13}$ cm$^{-2}$. This low doping concentration is not enough to shift the Fermi level of graphene to half of the incident photon energy (the photon energy corresponding to 1620 nm is about 0.77 eV and the half photon energy is about 0.38 eV). Therefore, the $\chi^{(3)}$ of graphene and the corresponding THG intensity does not change significantly after molecular adsorption.[4,5]

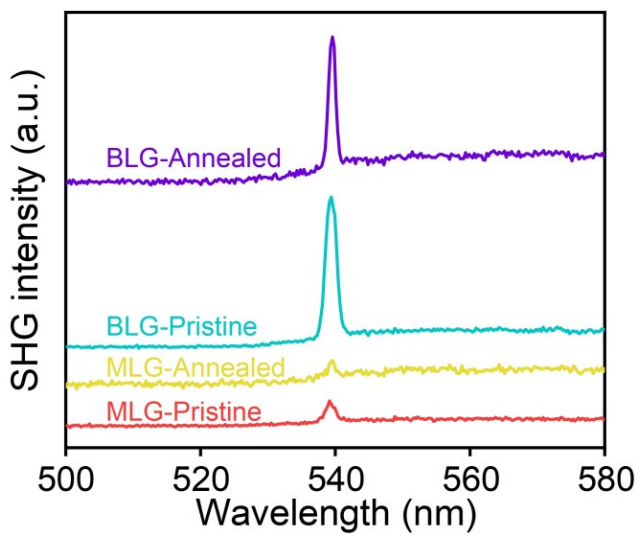

**Figure S4** THG intensities of MLG and BLG before and after molecular adsorption.

## 5. Calculation of the parallel and perpendicular components of the SHG intensities

The polarization-dependent SHG intensities could be analyzed from the second-order nonlinear susceptibility tensor **d**. We describe the SHG polarization **P** from the matrix calculation of **P** = **dE** in the $x-y$ coordinate defined in Figure 3a. The pump laser is incident along the $z$-axis. The **d** for a crystal of class $C_{3v}$ is[6]

$$\mathbf{d} = \begin{bmatrix} 0 & 0 & 0 & 0 & d_{31} & -d_{22} \\ -d_{22} & d_{22} & 0 & d_{31} & 0 & 0 \\ d_{31} & d_{31} & d_{33} & 0 & 0 & 0 \end{bmatrix} \quad (1)$$

By expressing **P** and **E** into their components in $x$, $y$, and $z$ directions, the calculation is given by[6]

$$\begin{bmatrix} P_x(2\omega) \\ P_y(2\omega) \\ P_z(2\omega) \end{bmatrix} = \begin{bmatrix} 0 & 0 & 0 & 0 & d_{31} & -d_{22} \\ -d_{22} & d_{22} & 0 & d_{31} & 0 & 0 \\ d_{31} & d_{31} & d_{33} & 0 & 0 & 0 \end{bmatrix} \begin{bmatrix} E_x(\omega)E_x(\omega) \\ E_y(\omega)E_y(\omega) \\ E_z(\omega)E_z(\omega) \\ 2E_y(\omega)E_z(\omega) \\ 2E_x(\omega)E_z(\omega) \\ 2E_x(\omega)E_y(\omega) \end{bmatrix} \quad (2)$$

For the electric field components ($E_x$, $E_y$, $E_z$) at the focus plane, because of the low numerical aperture of the used objective lens, $E_z$ could be neglected. Considering an electric field $E_0$ of the pump laser polarized along a direction with an angle $\theta$ to the $x$-axis, there are expressions of $E_x = E_0 \cos(\theta)$ and $E_y = E_0 \sin(\theta)$. The generated SHG polarization components ($P_x$, $P_y$, $P_z$) can be regarded as electric dipoles oscillating at the SHG frequency. The SHG signal radiated from $P_z$ can not be collected by the objective lens with a low numerical aperture. Therefore, only $P_x$ and $P_y$ contribute to the detected SHG intensity. The parallel and perpendicular components of the SHG intensities could be described as

$$I_{\parallel}(2\omega) \propto \left(P_x \cos(\theta) + P_y \sin(\theta)\right)^2 = \left(d_{22} \sin^3(\theta) - 3d_{22} \sin(\theta)\cos^2(\theta)\right)^2 = \left(d_{22} \sin(3\theta)\right)^2 \quad (3)$$

$$I_{\perp}(2\omega) \propto \left(P_x \sin(\theta) - P_y \cos(\theta)\right)^2 = \left(-d_{22} \cos^3(\theta) + 3d_{22} \cos(\theta)\sin^2(\theta)\right)^2 = \left(-d_{22} \cos(3\theta)\right)^2 \quad (4)$$

## 6. SHG in CVD-grown BLG

Figure S5a shows an optical microscope image of the sample transferred onto the same $SiO_2$/Si substrate via the PMMA-assisted wet transfer method. The hexagon region is AB-stacked BLG surrounded by MLG, as confirmed by the Raman spectroscopy (Figure S1). SHG measurement of the as-transferred sample is carried out first by spatially mapping, as shown in Figure S5b. Except for parts with wrinkled graphene, there is no observable SHG signal from both MLG and BLG. SHG from the wrinkled region could be attributed to the stacking-induced centrosymmetry breaking. Then, we anneal the transferred sample in a vacuum and expose it to an air ambient to involve the adsorptions of $O_2$ and $H_2O$ molecules. Interestingly, a significant SHG distribution is found in the spatial mapping image, as shown in Figure S5c. The polarization-resolved patterns of the emerging SHG intensity in Figure S5d are similar to that in Figure 4b, indicating the $C_{3v}$ point group after molecular adsorption. The solid lines are fitting curves, which are not in good agreement with the experimental results. It may be because the integrity of graphene lattice is destroyed during sample preparation and transfer, resulting in a weak SHG signal, which leads to the deviation between data and fitting. Although it deviates greatly from the fitting curves, it can still be seen that there are approximately six petals in each pattern. These results show that the emerging SHG induced by charge doping is also valid on the CVD-grown AB-stacked BLG samples.

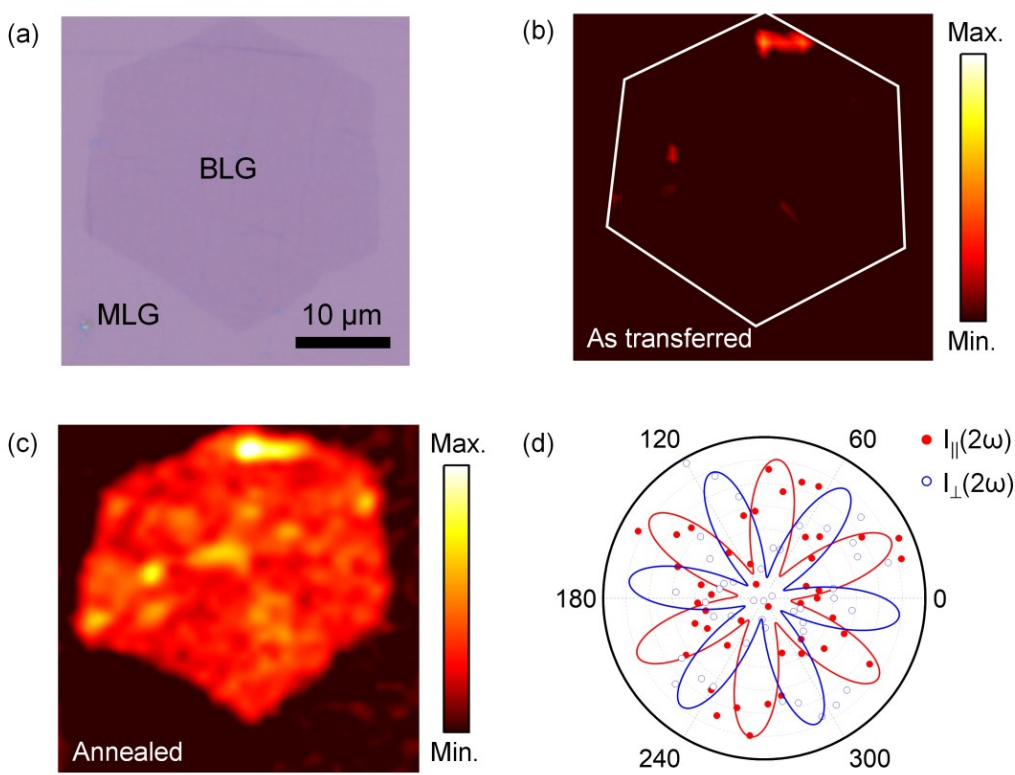

**Figure S5.** (a) Optical microscope image of a CVD-grown BLG sample transferred on the $SiO_2$/Si substrate. (b,c) SHG spatial mapping of the BLG sample (b) before and (c) after annealing. (d) Polarization-resolved patterns of SHG intensity from the CVD-grown BLG sample.

# References for Supporting Information